\documentclass[aps,prl,twocolumn]{revtex4}
\usepackage{amsfonts, amssymb, amsmath,graphicx}
\usepackage{subfigure}
\usepackage{multirow}
\usepackage{dcolumn}
\usepackage{bm}
\usepackage{color}

\definecolor{Nathanblue}{rgb}{0.,0.24,0.51}

\begin{document}

\title{Extracting the quantum metric tensor through periodic driving}

\author{Tomoki Ozawa}
\author{Nathan Goldman}
\affiliation{Center for Nonlinear Phenomena and Complex Systems, Universit\'e Libre de Bruxelles, CP 231, Campus Plaine, B-1050 Brussels, Belgium}%

\date{\today}

\newcommand{\tom}[1]{{\color{red} #1}}

\begin{abstract}
We propose a generic protocol to experimentally measure the quantum metric tensor, a fundamental geometric property of quantum states. Our method is based on the observation that the excitation rate of a quantum state directly relates to components of the quantum metric upon applying a proper time-periodic modulation. We discuss the applicability of this scheme to generic two-level systems, where the Hamiltonian's parameters can be externally tuned, and also to the context of Bloch bands associated with lattice systems. As an illustration, we extract the quantum metric of the multi-band Hofstadter model. Moreover, we demonstrate how this method can be used to directly probe the spread functional, a quantity which sets the lower bound on the spread of Wannier functions and signals phase transitions. Our proposal offers a universal probe for quantum geometry, which could be readily applied in a wide range of physical settings, ranging from circuit-QED systems to ultracold atomic gases.
\end{abstract}

\maketitle


The geometry of quantum states is at the heart of our understanding of diverse physical phenomena~\cite{Mead_review,Xiao:2010,Dalibard_gauge,Nakahara}, ranging from the Aharonov-Bohm effect~\cite{Aharonov,WuYang,Berry:1984} to the more recent topological states of matter~\cite{Thouless:1982,Hasan:2010,Qi:2011}. In non-relativistic quantum mechanics, the geometry of quantum states is typically associated with their parallel transport over the space spanned by the Hamiltonian's parameters~\cite{Nakahara,Simon1983}. Upon performing a closed loop in this parameter space, the state can acquire a finite geometric (Berry) phase~\cite{Berry:1984}, which can be attributed to the existence of a curvature~\cite{WuYang,Nakahara}. This so-called Berry curvature, whose effects were originally identified in the context of the anomalous Hall effect~\cite{Karplus}, played a crucial role in the development of topological band theory~\cite{Xiao:2010,Hasan:2010,Qi:2011}. More recently, the realization of engineered materials~\cite{Aidelsburger_review,Ozawa:2018, Cooper:2018} allowed for direct measurements of the Berry curvature, through state tomography~\cite{Flaschner:2016,Li:2016Science}, interferometry~\cite{Bloch_AB}, and transport~\cite{Wimmer:2017}.

Closely related to the Berry curvature is the quantum metric  tensor (or Fubini-Study metric), which is a distinct geometric property of energy eigenstates that reflects the ``distance" between different quantum states~\cite{Provost:1980, Anandan:1990,Marzari:1997}. The significance of the quantum metric was recently identified in a wide range of physical phenomena, including the conductivity in dissipative systems~\cite{Neupert:2013,Kolodrubetz:2013,Kolodrubetz:2017, Albert:2016, Bleu:2016, Ozawa:2018PRB}, orbital magnetism~\cite{Raoux:2015,Gao:2015,Piechon:2016,Combes:2016,Freimuth:2017}, the superfluid fraction~\cite{Julku:2016,Liang:2017,Iskin:2018}, quantum information~\cite{CamposVenuti:2007, Zanardi:2007, You:2007, Dey:2012}, entanglement and many-body properties~\cite{Ma:2010, Legner:2013, Roy:2014, Claassen:2015, Bauer:2016}, interference in Bloch states~\cite{Lim:2015PRA}, Lamb-shift-like energy shift in excitons~\cite{Srivastava:2015} and the mathematical construction of maximally-localized Wannier functions in crystals~\cite{Marzari:1997, Souza:2000, Resta:2005, Thonhauser:2006PRB, Marzari:2012}. Despite the importance of the quantum metric in these various contexts, one still lacks a direct experimental measurement of this geometric object.

In this paper, we propose a versatile experimental scheme to extract the components of the quantum metric, which is applicable to any parameter space. Our proposal consists in initially preparing the system in an eigenstate of a given Hamiltonian, and then monitoring the excitation rate upon modulating the Hamiltonian periodically in time. As shown below, the excitation rate is shown to be proportional to the quantum metric for suitable choices of the periodic driving. This protocol is inspired by previous works connecting absorption properties with quantum geometry~\cite{Souza_2008,Tran:2017,Juan:2017,Tran:2018}, and is further motivated by the recent development of excitation-rates measurements in shaken atomic gases~\cite{Aidelsburger:2014,Reitter2017,Flaschner_2018}. 

We first discuss the case where the Hamiltonian depends on a set of parameters, which can be externally tuned and modulated in time. We consider a generic two-level system as an example, based on which we numerically demonstrate the applicability of our method. We then extend the scheme to the case of lattice systems, where the parameters of the Bloch Hamiltonian are crystal momenta defined in the Brillouin zone. We numerically validate our method using the multi-band Harper-Hofstadter model~\cite{Harper:1955,Hofstadter:1976} and the two-band Haldane model~\cite{Haldane:1998}. Finally, we show how to measure the  ``spread functional", i.e.~the trace of the quantum metric averaged over the Brillouin zone, which is related to the localization of Wannier functions~\cite{Marzari:1997}, and which was shown to signal phase transitions~\cite{Thonhauser:2006PRB}. The protocols and examples discussed in this work are of direct experimental relevance to synthetic quantum systems such as circuit-QED setups~\cite{Roushan:2014Nature} and ultracold atomic gases~\cite{Lin:2009Nature,Jotzu:2015, Aidelsburger:2013, Miyake:2013, Aidelsburger:2014, Kennedy:2015, Flaschner:2016, Tarnowski:2017PRL}.

\textit{Introducing the quantum metric.} Let us start by considering a quantum state $|\boldsymbol\lambda\rangle$, which depends on a set of dimensionless parameters $\boldsymbol\lambda = (\lambda_1, \lambda_2, \cdots, \lambda_N)$, where $N$ is the dimension of the parameter space. The quantum metric $g_{\mu\nu}(\boldsymbol\lambda)$ defined in this parameter space originates from the so-called quantum geometric tensor~\cite{Kolodrubetz:2017}
\begin{align}
	\chi_{\mu \nu} (\boldsymbol\lambda)
	\equiv
	\langle \partial_{\mu} \boldsymbol\lambda| \left(1 - |\boldsymbol\lambda\rangle \langle \boldsymbol\lambda| \right) |\partial_{\nu}\boldsymbol\lambda\rangle ,
\end{align}
which defines a gauge-invariant quantity associated with $\vert\boldsymbol\lambda\rangle$. The  tensor $\chi_{\mu \nu}$ can take complex values:~the real part defines the quantum metric $g_{\mu\nu}(\boldsymbol\lambda) \!\equiv\! \mathrm{Re}\left[\chi_{\mu \nu} (\boldsymbol\lambda)\right]$, while the imaginary part $\mathrm{Im}\left[\chi_{\mu \nu} (\boldsymbol\lambda)\right] \!=\!-\Omega_{\mu \nu}(\boldsymbol\lambda)/2$ is related to the Berry curvature $\Omega_{\mu \nu}(\boldsymbol\lambda)$. The quantum metric is symmetric, $g_{\mu \nu}(\boldsymbol\lambda) \!=\! g_{\nu \mu}(\boldsymbol\lambda)$, and it provides a distance~\cite{Kolodrubetz:2017} between nearby states $\vert \boldsymbol\lambda\rangle$ and $\vert \boldsymbol\lambda + \text{d}\boldsymbol\lambda\rangle$.

\textit{General formalism in parameter space.}
We now consider a Hamiltonian $\hat{H}(\boldsymbol\lambda)$ that depends on a set of dimensionless parameters $\boldsymbol\lambda$.
The energy eigenstates also depend on $\boldsymbol\lambda$, and the quantum metric can be defined for each of them.
We assume that the system is initially prepared at $\boldsymbol\lambda\!=\!\boldsymbol\lambda^0$, and in order to detect the quantum metric at that point, we modulate one parameter ($\lambda_1$) as
\begin{align}
	\lambda_1(t) = \lambda_{1}^0 + 2(E/\hbar \omega) \cos (\omega t).
\end{align}
Assuming that the amplitude of the modulation is small, $(E/\hbar \omega) \ll 1$, we Taylor-expand the Hamiltonian and obtain
\begin{align}
	\hat{H}[\boldsymbol\lambda (t)]
	=
	\hat{H}(\boldsymbol\lambda^0)
	+
	\partial_{\lambda_1} \hat{H}(\boldsymbol\lambda^0) 2 (E/\hbar \omega) \cos (\omega t).\label{ham_general_pert}
\end{align}
At time $t \!=\! 0$, the system is assumed to be in the lowest-energy eigenstate $|i\rangle$ of the Hamiltonian, $\hat{H}(\boldsymbol\lambda^0)|i\rangle \!=\! \epsilon_i |i\rangle$. Then, according to time-dependent perturbation theory~\cite{LandauBook}, the probability of observing the system in another eigenstate $|f\rangle$ at time $t$ is given by~\footnote{  In Eq.~\eqref{excitations}, we assumed that: (a) the observation time is sufficiently long $t\!\gg1/\omega$ so as to neglect anti-resonant contributions; (b) the excited fraction remains small over the observation time, $t \ll \hbar^2 \omega / \left\{ E |\langle f| \partial_{\lambda_1} \hat{H}(\boldsymbol\lambda^0)|i\rangle|\right\}$.}
\begin{align}
	n_f(\omega,t)
	\!=\!\frac{2\pi t}{\hbar}\,\left|
	\frac{E}{\hbar \omega}
	\langle f| \partial_{\lambda_1} \hat{H}(\boldsymbol\lambda^0)|i\rangle
	\right|^2\!\delta (\epsilon_f \!-\! \epsilon_i \!-\! \hbar \omega).\label{excitations}
\end{align}
In the following, we are interested in the total excitation rate, which is obtained by summing Eq.~\eqref{excitations} over all possible final states: $\Gamma(\omega)\!\equiv\!(1/t) \sum_f n_f(\omega, t)$. Inspired by Ref.~\cite{Tran:2017}, we now introduce the \emph{integrated rate}:
\begin{align}
	\Gamma^\mathrm{int}
	&\equiv
	\int d\omega\, \Gamma(\omega)
	=
	\frac{2\pi E^2}{\hbar^2}
	\sum_f
	\frac{\left| \langle f| \partial_{\lambda_1} \hat{H}(\boldsymbol\lambda^0)|i\rangle \right|^2}
	{(\epsilon_f - \epsilon_i)^2}.
\end{align}
Using the identity $ \langle f| \partial_{\lambda_1} \hat{H}(\boldsymbol\lambda^0)|i\rangle \!=\! -(\epsilon_f - \epsilon_i)\langle f|\partial_{\lambda_1}i\rangle$,
one eventually obtains the relation between the integrated rate and the \emph{diagonal} components of the quantum metric:
\begin{align}
	\Gamma^\mathrm{int}
	&=
	\frac{2\pi E^2 }{\hbar^2}
	\sum_f
	\left| \langle f|\partial_{\lambda_1}i\rangle \right|^2
	=
	\frac{2\pi E^2}{\hbar^2}
	g_{\lambda_1 \lambda_1} (\boldsymbol\lambda^0).\label{diag_gen}
\end{align}
Similarly, one can relate the \emph{off-diagonal} components of the quantum metric to the excitation rate $\Gamma$, by combining two sets of measurements, as we now explain.
Consider modulating two parameters $\lambda_1$ and $\lambda_2$ as
\begin{align}
	\lambda_1(t) = \lambda_{1}^0 + 2(E/\hbar \omega) \cos (\omega t), \notag \\
	\lambda_2(t) = \lambda_{2}^0 \pm 2(E/\hbar \omega) \cos (\omega t).\label{lambda2}
\end{align}
The probability of finding the system in a state $|f\rangle$ is
\begin{align}
	n_f^\pm(\omega, t)
	&=
	\frac{E^2}{(\hbar \omega)^2}
	\left|
	\langle f| \partial_{\lambda_1} \hat{H}(\boldsymbol\lambda^0) \pm \partial_{\lambda_2} \hat{H}(\boldsymbol\lambda^0) |i\rangle
	\right|^2
	\notag \\
	&\phantom{=}
	\times 2\pi \frac{t}{\hbar}\delta (\epsilon_f - \epsilon_i - \hbar \omega),
\end{align}
where $\pm$ refers to the sign in Eq.~\eqref{lambda2}.
In this scenario, the integrated rate becomes
\begin{align}
	\Gamma^{\mathrm{int}}_{\pm}
	&\equiv
	(1/t)\int d\omega\, \sum_f n_f^{\pm}(\omega, t)
	\notag \\
	&=
	\frac{2\pi E^2}{\hbar^2}
	\left(
	g_{\lambda_1 \lambda_1}(\boldsymbol\lambda^0)
	\pm
	2 g_{\lambda_1 \lambda_2}(\boldsymbol\lambda^0)
	+
	g_{\lambda_2 \lambda_2}(\boldsymbol\lambda^0)
	\right).\notag
\end{align}
Then, taking their difference, we obtain
\begin{align}
	\Delta \Gamma^{\text{int}}=\Gamma^{\mathrm{int}}_{+}-\Gamma^{\mathrm{int}}_{-}
	=
	\frac{8\pi E^2}{\hbar^2}
	g_{\lambda_1 \lambda_2}(\boldsymbol\lambda^0),\label{eq:DIR}
\end{align}
which relates the differential integrated rate to \emph{off-diagonal} elements of the quantum metric~\footnote{Had we replaced $\lambda_2(t)$ in Eq.~\eqref{lambda2} by the modulation $\lambda_2(t)\!=\! \lambda_{2}^0 \pm 2(E/\hbar \omega) \sin (\omega t)$, this differential measurement would have provided the Berry curvature $\Omega_{\lambda_1 \lambda_2}(\boldsymbol\lambda^0)$, similarly to Ref.~\cite{Tran:2017}.}.

\textit{Application to two-level systems.}
We apply the proposed scheme to a generic two-level system, whose Hamiltonian can be parametrized by two parameters $(\theta, \phi)$,
\begin{align}
	\hat{H}(\theta, \phi)
	=
	H_0
	\begin{pmatrix}
	\cos \theta & \sin \theta e^{-i\phi} \\
	\sin \theta e^{i\phi} & - \cos \theta
	\end{pmatrix} ; \label{hambloch}
\end{align}
see Refs.\cite{Roushan:2014Nature,Dalibard_gauge} for  realizations where $(\theta, \phi)$ can be tuned. The eigenenergies of the system are $\pm H_0$,  and the quantum metric is found to be independent of the level~\cite{Kolodrubetz:2017}
\begin{align}
	g_{\theta \theta} &= 1/4, &
	g_{\phi \phi} &= (\sin^2 \theta)/4, & 
	g_{\theta \phi} &= 0. \label{fsbloch}
\end{align}

\begin{figure}[htbp]
\begin{center}
\includegraphics[width= 0.37 \textwidth]{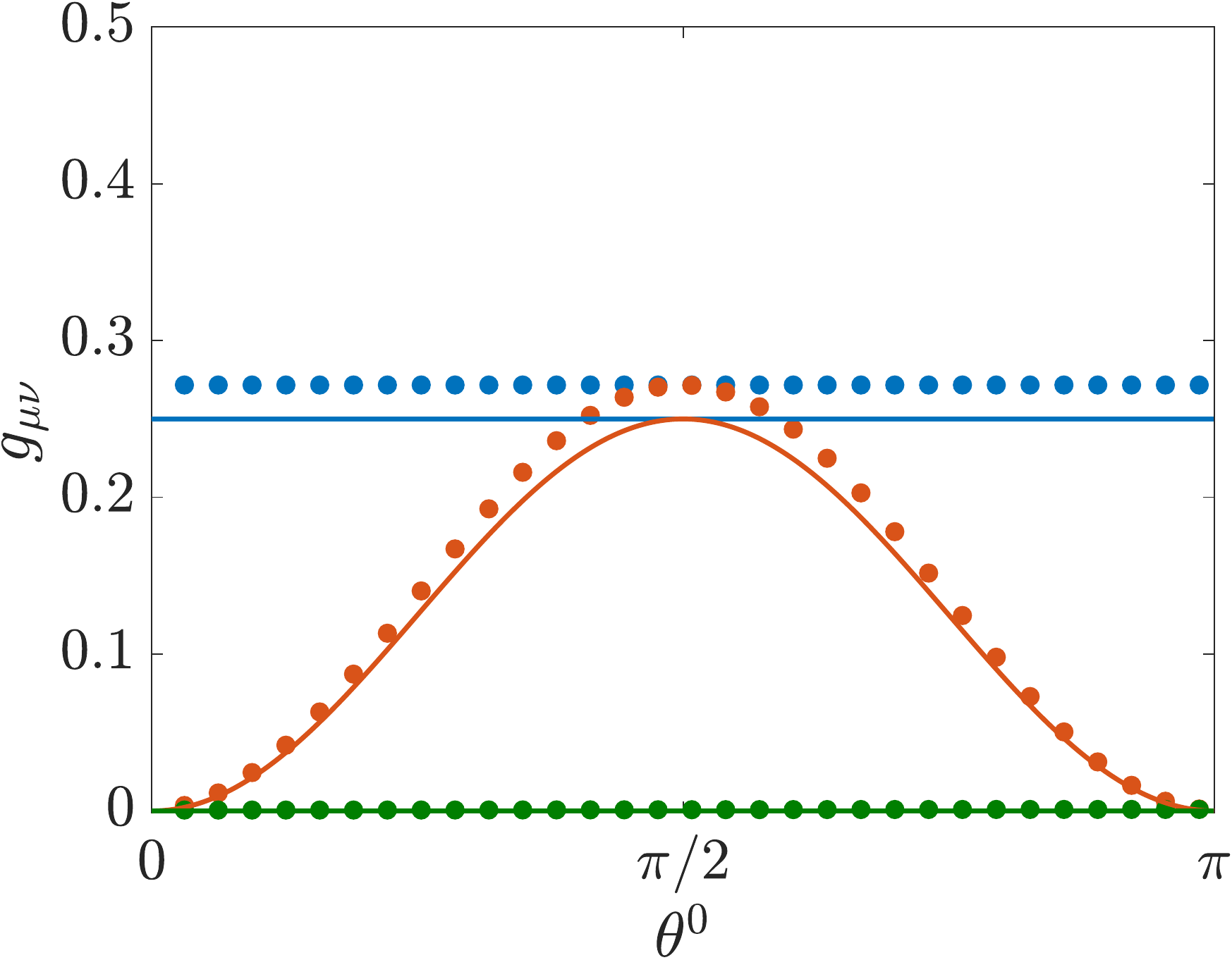}
\caption{Extraction of the quantum metric of the two-level system (\ref{hambloch}), using the protocol [Eqs.~\eqref{diag_gen} and \eqref{eq:DIR}]. The numerically extracted values of $g_{\theta\theta}$ (top blue dots), $g_{\phi\phi}$ (middle orange dots), and $g_{\theta\phi}$ (bottom green dots) are plotted together with their ideal values [Eq.~\eqref{fsbloch}] in full lines. The full-time-dynamics simulations are performed with the drive amplitude $E \!=\! 0.01H_0$, up to the observation time $t \!=\! 20\hbar/H_0$. The required integral over $\omega$ has been calculated using the range $\hbar\omega\!\in\![0.5H_0 , 3.5H_0]$, and the discrete step $\hbar\delta \omega \!=\! 0.05 H_0$. In all simulations, we set the initial condition $\phi^0 \!=\! \pi/4$.}
\label{bloch_extracted}
\end{center}
\end{figure}

We now numerically explore the validity of the proposed scheme [Eqs.~\eqref{diag_gen} and \eqref{eq:DIR}]. In order to extract $g_{\theta\theta}$, we prepare a state in the lower-energy eigenstate of the Hamiltonian~$\hat{H}(\theta^0, \phi^0)$, and then modulate $\theta$ in time according to
$ \theta(t) \!=\! \theta^0 \!+\! 2 (E/\hbar \omega) \cos(\omega t)$. We simulate the full time-evolution of the driven system and measure the population in the upper state as a function of time; repeating this procedure for many values of the drive frequency $\omega$ allows one to evaluate $g_{\theta\theta}$ through Eq.~\eqref{diag_gen}.
Similarly, we modulate $\phi(t)$ to extract $g_{\phi\phi}$, and finally modulate both $\theta(t)$ and $\phi(t)$ to extract $g_{\theta\phi}$ [Eqs.~\eqref{lambda2},\eqref{eq:DIR}]. The results presented in Fig.~\ref{bloch_extracted}, which displays the components of the extracted quantum metric as a function of the initial condition $\theta^0$ (for a fixed $\phi^0$), show very good agreement between the ideal values (solid lines) and those extracted from numerical simulations (dots). Slight deviations are attributed to the finite interval $\hbar \omega\!\in\![0.5H_0 , 3.5H_0]$ used to calculate the integrated rates, and/or to finite observation times.

\textit{Application to lattice systems.} We now extend the method to the case of Bloch states, where the relevant parameter space is spanned by crystal momentum. Since the latter quantity cannot be directly modulated in time, using an external drive, we now  build on the proposal~\cite{Tran:2017}, where a circular shaking of the lattice was shown to reveal the Berry curvature and Chern number of a populated Bloch band. To probe the quantum metric, we propose to shake the lattice \emph{linearly} as~\footnote{In a moving frame~\cite{Tran:2017}, the Hamiltonian in Eq.~\eqref{xshaking} becomes translationally-invariant and can be expressed as $\mathcal{H}_x (\mathbf{k};t)\!\approx\!\hat{H}_\mathrm{lattice}(\mathbf{k})\!+\!(2E/\hbar \omega)\sin(\omega t)\partial_{k_x}\hat{H}_\mathrm{lattice}(\mathbf{k})$, in direct analogy with Eq.~\eqref{ham_general_pert}.}
\begin{align}
	\hat{H}_x (t) = \hat{H}_\mathrm{lattice} + 2E \cos (\omega t) \hat{x}, \label{xshaking}
\end{align}
where $\hat{H}_\mathrm{lattice}$ is the lattice Hamiltonian of interest and $\hat{x}$ is the position operator. For the sake of clarity, we first assume that the system is initially prepared in a given Bloch state, $|\psi_\mathrm{ini}\rangle\!=\!e^{i\mathbf{k}^0 \cdot \mathbf{r}}|u_n(\mathbf{k}^0)\rangle$, where $\mathbf{k}^0$ is the corresponding quasi-momentum and $n$ is the band index. Considering an initial state in the lowest band $\epsilon_1(\mathbf{k})$, the excitation rate upon linear drive [Eq.~\eqref{xshaking}] is given by
\begin{align}
	\Gamma_{\hat{x}} (\omega)
	=
	&\sum_{u_n(\mathbf{k}) \neq u_1(\mathbf{k}^0)}|\langle u_n(\mathbf{k}) | e^{-i\mathbf{k}\cdot \mathbf{r}} \hat{x} e^{i\mathbf{k}^0\cdot \mathbf{r}}|u_1(\mathbf{k}^0)\rangle|^2
	\notag \\
	&\phantom{=}
	\qquad \quad \times \frac{2\pi}{\hbar}E^2 \delta (\epsilon_n(\mathbf{k}) - \epsilon_1(\mathbf{k}^0) - \hbar \omega).
\end{align}
Noting that the position operator is diagonal and acts as a derivative in $k$-space~\cite{Karplus},
$ \langle u_n(\mathbf{k}) | e^{-i\mathbf{k}\cdot \mathbf{r}} \hat{x} e^{i\mathbf{k}^0\cdot \mathbf{r}} |u_1(\mathbf{k}^0)\rangle \!=\! i\delta_{\mathbf{k},\mathbf{k}^0} \langle u_n(\mathbf{k}) |\partial_{k_x} u_1(\mathbf{k})\rangle$, 
and integrating the rate $\Gamma_{\hat{x}} (\omega)$ over $\omega$, we obtain
\begin{align}
	\Gamma_{\hat{x}}^{\mathrm{int}}
	&\equiv
	\int d\omega\, \Gamma_{\hat{x}} (\omega)
	=
	\frac{2\pi E^2}{\hbar^2} \sum_{n \neq 1}|\langle u_n(\mathbf{k}^0) | \partial_{k_x} u_1(\mathbf{k}^0)\rangle|^2
	\notag \\
	&=
	\frac{2\pi E^2}{\hbar^2} g_{xx}^1(\mathbf{k}^0), \label{estimatexx}
\end{align}
where $g_{xx}^1$ is the $xx$-component of the quantum metric associated with the lowest band.
Similarly, the $yy$-component is related to the integrated rate upon linear shaking along the $y$-direction.

In order to obtain the $xy$-component of the quantum geometric tensor, we apply the linear shaking along the diagonal directions $\hat{x} \pm \hat{y}$, 
\begin{align}
	\hat{H}_{x\pm y} (t) = \hat{H}_\mathrm{lattice} + 2E \cos (\omega t) (\hat{x} \pm \hat{y}) , \label{diag_shake}
\end{align}
which results in the integrated rate
\begin{align}
	\Gamma_{\hat{x}\pm\hat{y}}^{ \mathrm{int}}
	=
	\frac{2\pi}{\hbar^2}E^2 \left( g_{xx}^1(\mathbf{k}^0) \pm 2g_{xy}^1(\mathbf{k}^0) + g_{yy}^1(\mathbf{k}^0) \right).
\end{align}
Upon taking their difference, we find the relation between $g_{xy}^1$ and the differential integrated rate
\begin{align}
	\Delta \Gamma^{\text{int}} \equiv \Gamma_{\hat{x} + \hat{y}}^{ \mathrm{int}} - \Gamma_{\hat{x}-\hat{y}}^{ \mathrm{int}}
	=
	\frac{8\pi}{\hbar^2}E^2 g_{xy}^1(\mathbf{k}^0).\label{eq:DIR_lattice}
\end{align}
We stress that the relation~\eqref{eq:DIR_lattice} only involves linearly-polarized modulations [Eq.~\eqref{diag_shake}], which is in contrast to analogous Berry-curvature measurements obtained from circularly-polarized modulations~\cite{Tran:2017}.
We also note that
\begin{align}
	\Sigma \Gamma^{\text{int}} \equiv \Gamma_{\hat{x} + \hat{y}}^{ \mathrm{int}} + \Gamma_{\hat{x}-\hat{y}}^{ \mathrm{int}}
	=
	\frac{4\pi}{\hbar^2}E^2 \mathrm{Tr}\left[g_{\mu\nu}^1(\mathbf{k}^0) \right], \label{eq:DIR_lattice_trace}
\end{align}
offers a direct probe for the quantum metric trace.

So far, we have assumed that the initial state corresponds to a single eigenstate of the lattice Hamiltonian. However, the results above directly generalize to the case where the initial state is made of a superposition of states in the lowest band,
$ |\psi_\mathrm{ini}\rangle \!=\! \sum_\mathbf{k} c(\mathbf{k}) e^{i\mathbf{k}\cdot\mathbf{r}}|u_1 (\mathbf{k})\rangle$. For instance, upon adding a linear shaking along $x$ [Eq.~\eqref{xshaking}], the integrated rate in Eq.~\eqref{estimatexx} is now given by the weighted average of the quantum metric:
\begin{align}
	\Gamma_{\hat{x}}^{\mathrm{int}}
	=
	\frac{2\pi}{\hbar}E^2 \sum_{\mathbf{k}}|c(\mathbf{k})|^2 g_{xx}^1 (\mathbf{k}). \label{extractingaverage}
\end{align}
A similar relation holds for non-interacting fermions  (partially) filling the lowest band, in which case the weight $|c(\mathbf{k})|^2$ in Eq.~\eqref{extractingaverage} should be replaced by the density of fermions $\rho (\mathbf{k})$ in this band.

\begin{figure}[htbp]
\begin{center}
\includegraphics[width= 0.37 \textwidth]{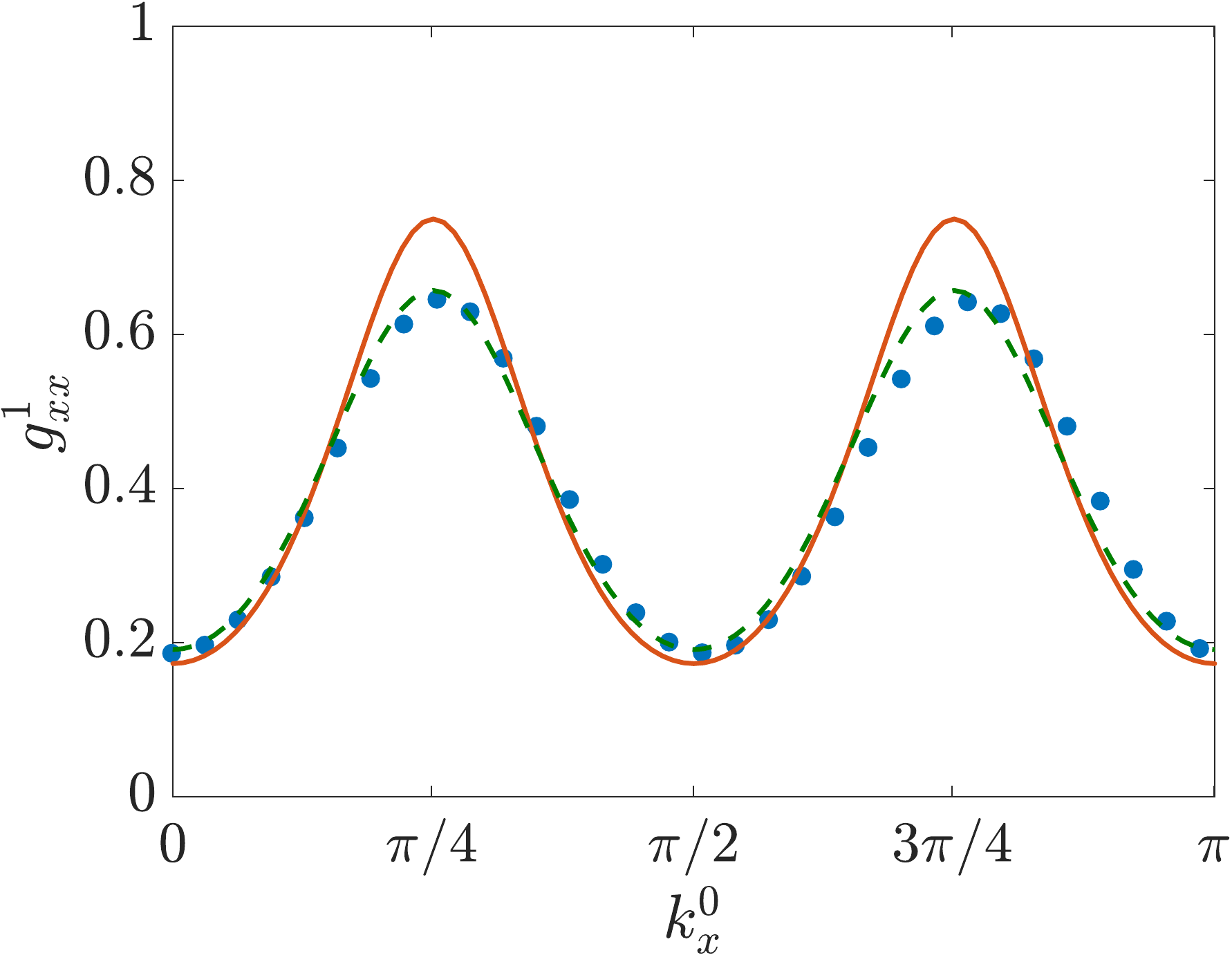}
\caption{The quantum metric $g_{xx}^1$ associated with the lowest band of the Harper-Hofstadter model, with flux $\alpha \!=\! \pi/2$ per plaquette, as extracted by linearly shaking the lattice along $x$ [Eq.~\eqref{estimatexx}]. Dots are extracted values of $g_{xx}^1$, comparing numerical dynamics with Eq.~\eqref{estimatexx}, the orange line is the ideal value of $g_{xx}^1$, and the green line is the weighted average $\sum_{\mathbf{k}}|c(\mathbf{k})|^2 g_{xx}^1(\mathbf{k})$, where $|c(\mathbf{k})|^2$ reflects the Gaussian profile of the initial wavepacket. The simulations are done on a lattice of size 45 $\times$ 45, up to the final time of $t \!=\! 10 \hbar/|J|$, and the drive strength is $E \!=\! 0.01|J|$, where $J$ is the hopping amplitude. We integrate the rates over $\omega$, using the range $\hbar\omega\!\in\![0.5|J| , 5.5 |J|]$, and a discrete step $\hbar\delta \omega \!=\! 0.05|J|$.}
\label{hh_extracted}
\end{center}
\end{figure}

\textit{Example: Harper-Hofstadter model.}
We numerically explore the validity of this method by considering the Harper-Hofstadter model~\cite{Harper:1955, Hofstadter:1976}, and setting the flux per plaquette $\alpha \!=\! \pi/2$ as in recent cold-atom experiments~\cite{Aidelsburger:2013,Aidelsburger:2014}. In order to extract $g_{xx}^1(\mathbf{k}^0)$, around a given quasi-momentum, we initially prepare a Gaussian wavepacket of the form $\psi_{\mathrm{ini}} (\mathbf{x}; \mathbf{k}^0) \!\propto\! e^{i k_x^0 x + ik_y^0 y -(x^2 + y^2)/2\sigma^2}$, which we then project unto the lowest band; experimentally, this would correspond to adiabatically loading a wave-packet into the lowest (Floquet) band~\cite{Aidelsburger:2014}, and setting $\mathbf{k}^0$ through Bloch oscillation. Here, $\sigma$ denotes the spread of the wavepacket in real space, and the lattice spacing is unity. As discussed above [Eq.~\eqref{extractingaverage}], measuring the integrated rate $ \Gamma_{\hat{x}}^{\mathrm{int}}$ upon linear shaking along $x$ then provides an estimation of the quantum metric $g_{xx}^1(\mathbf{k}^0)$, with a resolution that improves as one reduces the  spread $1/\sigma$ in $k$-space. We have implemented this protocol through a full-time-dynamics simulation of the driven Harper-Hofstadter model. The results are shown in Fig.~\ref{hh_extracted}, which displays the quantum metric $g_{xx}^1(\mathbf{k}^0)$  extracted from the simulated dynamics and using the relation in Eq.~\eqref{estimatexx}. Here we fixed $k_y^0 \!=\! \pi/4$, and measured $g_{xx}^1$ for various $k_x^0$; see the blue dots in Fig.~\ref{hh_extracted}. These numerical results are compared to the ideal values of $g_{xx}^1(\mathbf{k})$ (solid orange line), together with the weighted-average of $g_{xx}^1(\mathbf{k})$ using the Gaussian profile (dashed green line). The perfect agreement between the numerical simulation and the weighted-average prediction~\eqref{extractingaverage} validates the approach. As discussed above, the extracted values of $g_{xx}^1$ can be made closer to the ideal theory prediction  [Eq.~\eqref{estimatexx}] by improving the localization of the wavepacket in momentum space.

\textit{Spread functional in the Haldane model.}
As a final example, we show how our protocol can be used to directly extract the spread functional~\cite{Marzari:1997}, which characterizes the localization of Wannier functions in a given Bloch band. First, recall that the quadratic spread of  Wannier functions is lower-bounded by the trace of the quantum metric according to~\cite{Marzari:1997, Resta:2005, Thonhauser:2006PRB, Marzari:2012}
\begin{align}
	\langle r^2 \rangle -\langle r \rangle^2 \ge \overline{\mathrm{Tr}[g_{\mu \nu}(\mathbf{k})]} \equiv \Omega_\mathrm{I},\label{Wannier_test}
\end{align}
where $\overline{\mathrm{Tr}[g_{\mu \nu}(\mathbf{k})]}$ is the average of $g_{xx} + g_{yy}$ over the Brillouin zone. As suggested in Eq.~\eqref{eq:DIR_lattice_trace}, the quantity $\Omega_\mathrm{I}$ could be directly extracted from our protocol upon filling a given Bloch band uniformly over the Brillouin zone. Interestingly, this could offer a first experimental detection of this abstract quantity, which played such a crucial role in constructing maximally-localized Wannier functions~\cite{Marzari:1997, Souza:2000, Resta:2005, Thonhauser:2006PRB, Marzari:2012}.

We demonstrate this scheme by considering the Haldane lattice model~\cite{Haldane:1998,Jotzu:2015}, which is characterized by the staggered potential strength $M$ and the second-neighbor hopping matrix elements $\pm iJ^\prime $. This model exhibits a topological phase transition at $M\!=\!3\sqrt{3}J^\prime$, which is associated with a change in the Chern number. In order to access the components $g_{xx}^1$ and $g_{yy}^1$, we numerically shake the lattice along the $\hat{x}$ and $\hat{y}$ directions, respectively. Furthermore, the average of these components over the Brillouin zone are readily obtained by preparing an initial state that is evenly distributed over the entire lowest band. In our numerical simulations, this is achieved by considering a state in a single lattice site and projecting the latter unto the desired band; see~\cite{Sup} for details. In experiments, the uniform filling is directly achieved through Pauli statistics.

\begin{figure}[htbp]
\begin{center}
\includegraphics[width= 0.37 \textwidth]{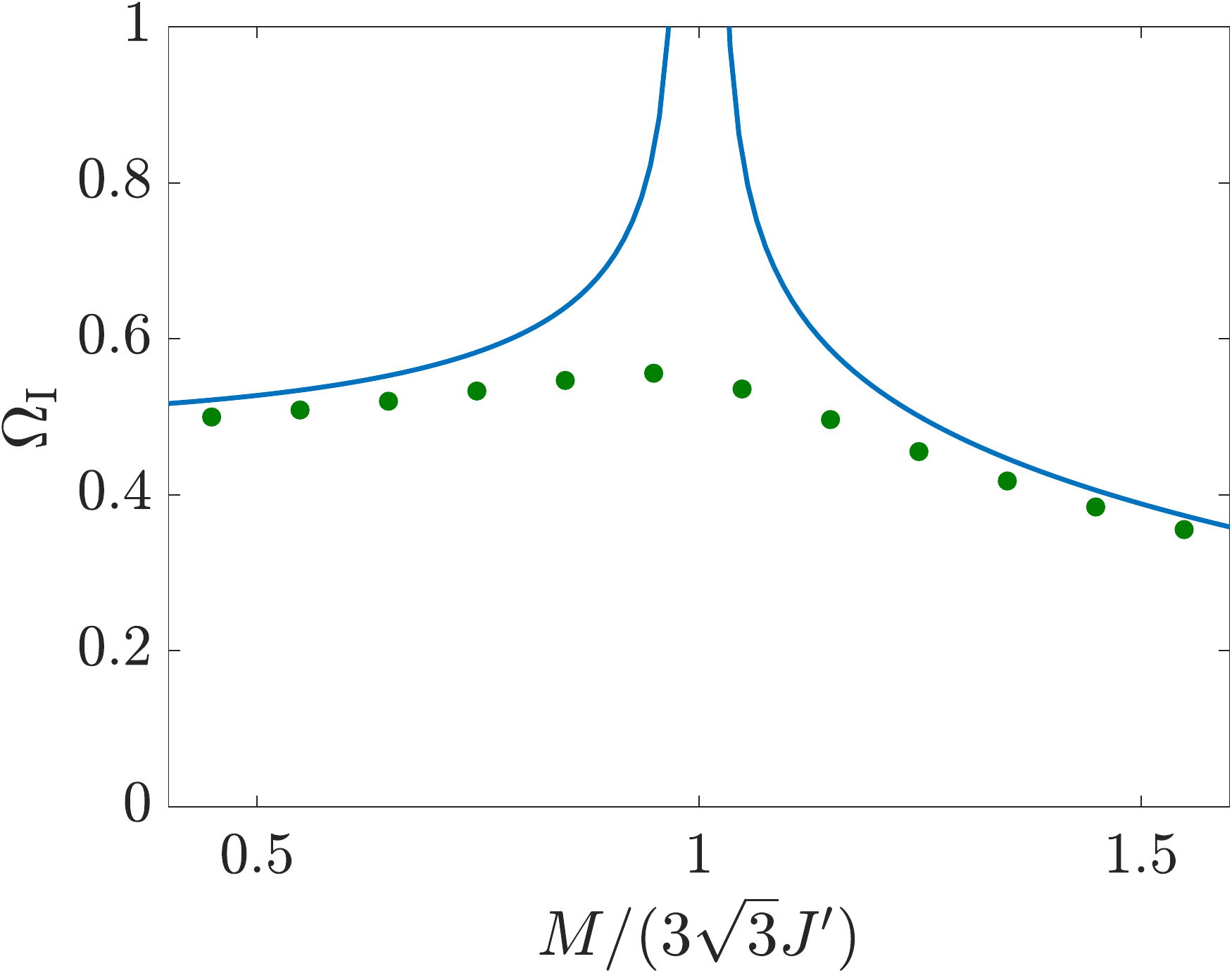}
\caption{Values of the spread functional $\Omega_\mathrm{I}$, as extracted from the protocol applied to the shaken Haldane model (dots). The solid line shows the ideal values of $\Omega_\mathrm{I}$. The simulations are done on a lattice with 45 $\times$ 45 unit cell with $J^\prime \!=\! 0.1J$, where $J$ is the nearest-neighbor hopping amplitude. The amplitude of the modulation is $E \!=\! 0.01J$ and the simulation is performed up to the time of $t \!=\! 10\hbar/J$. We integrate the rates over $\omega$, using the range $\hbar\omega\!\in\![0.05J , 5.0 J]$,  and a discrete step $\hbar\delta \omega \!=\! 0.05 J$.}
\label{fig:localization}
\end{center}
\end{figure}

Figure~\ref{fig:localization} shows $\Omega_\mathrm{I}$ as extracted from our simulations of the shaken Haldane model, and compared with their ideal values (solid line).  As noted in Ref.~\cite{Thonhauser:2006PRB}, $\Omega_\mathrm{I}$ diverges logarithmically at the topological phase transition,  which signals the crossing through a metallic regime. Away from this phase transition, we find that the numerical values $\Omega_\mathrm{I}$ associated with our protocol well capture the behavior of the ideal spread functional. However, the divergent feature of this quantity at the transition is difficult to probe, which is due to the limitation of the method when dealing with arbitrarily small frequencies.

We stress that the relation between $\Omega_\mathrm{I}$ and the localization of Wannier functions [Eq.~\eqref{Wannier_test}] is only meaningful in the trivial regime ($M/3\sqrt{3}J^\prime > 1$); see Ref.~\cite{Thonhauser:2006PRB}. In this regime, the protocol described above could be used to experimentally measure the localization of Wannier states, which could then be used to validate the numerical construction of Wannier functions for systems of interest.


\paragraph{Acknowledgments} We are grateful to helpful discussions with M. Di Liberto, P. Hauke, M. Kolodrubetz, B. Mera, G. Palumbo, G. Salerno, and D.-T. Tran. This work was supported by the FRS-FNRS (Belgium) and the ERC Starting Grant TopoCold.

\clearpage

\begin{widetext}

\begin{center}
\begin{large}
{\bf Supplemental Material for \\``Extracting the quantum metric tensor through periodic driving"}
\end{large}
\end{center}

\end{widetext}

\section{Uniform filling through single-site occupations}
In order to measure the average of the quantum metric of a lowest band over the Brillouin zone, we need to prepare an initial state that is uniformly distributed over the Brillouin zone. The situation can be straightforwardly realized by considering non-interacting fermions (which can automatically fill the lowest band according to Pauli principle). An equivalent situation, which is numerically much less costly to simulate, can be realized by preparing an initial state that is localized on a single site in real-space. In this Supplemental Material, we show the latter approach indeed gives rise to the average of the quantum metric over the Brillouin zone.

For concreteness, we consider a tight-binding model with two lattice sites per unit cell, such as the Haldane model analyzed in the main text. Extension to more general cases is straightforward. Let us denote the Bloch state wavefunction in band $n$ with crystal momentum $\mathbf{k}$ as $e^{i\mathbf{k}\cdot \mathbf{r}}|u_n (\mathbf{k})\rangle$. Here, the cell-periodic part of the Bloch state $|u_n (\mathbf{k})\rangle$ is a vector with two components. A state that is localized at the origin of the system, which we assume is in one of the two sublattices (which we call A sublattice), can be written as
\begin{align}
	|\mathrm{A}\rangle
	=
	\frac{1}{\sqrt{N_\mathrm{cell}}}
	\sum_\mathbf{k} e^{i\mathbf{k}\cdot \mathbf{r}}
	\begin{pmatrix}
	1 \\ 0
	\end{pmatrix},
\end{align}
where $N_\mathrm{cell}$ is the number of unit cells in the system.
The cell-periodic part of the Bloch states $|u_1 (\mathbf{k})\rangle$ and $|u_2 (\mathbf{k})\rangle$ form a complete set of basis for a given value of $\mathbf{k}$. This implies that, by writing $|e_A\rangle \equiv (1,0)^T$ and $|e_B\rangle \equiv (0,1)^T$, there exists a unitary matrix
\begin{align}
	U(\mathbf{k})
	=
	\begin{pmatrix}
	a_1 (\mathbf{k}) & a_2 (\mathbf{k}) \\
	b_1 (\mathbf{k}) & b_2 (\mathbf{k})
	\end{pmatrix}, \label{unitarytransformation}
\end{align}
for each value of $\mathbf{k}$, which satisfies
\begin{align}
	\begin{pmatrix}
	|e_A\rangle \\ |e_B\rangle
	\end{pmatrix}
	=
	U(\mathbf{k})
	\begin{pmatrix}
	|u_1 (\mathbf{k})\rangle \\ |u_2 (\mathbf{k})\rangle
	\end{pmatrix}.
\end{align}
Using the components of $U(\mathbf{k})$, we can expand the state $|\mathrm{A}\rangle$ as
\begin{align}
	|\mathrm{A}\rangle
	=
	\frac{1}{\sqrt{N_\mathrm{cell}}}
	\sum_\mathbf{k} e^{i\mathbf{k}\cdot \mathbf{r}}
	\left(
	a_1 (\mathbf{k})|u_1(\mathbf{k})\rangle
	+
	a_2 (\mathbf{k})|u_2(\mathbf{k})\rangle
	\right).
\end{align}
Similarly, the state which is localized in the other (B) sublattice can be written as
\begin{align}
	|\mathrm{B}\rangle
	&=
	\frac{1}{\sqrt{N_\mathrm{cell}}}
	\sum_\mathbf{k} e^{i\mathbf{k}\cdot \mathbf{r}}
	\begin{pmatrix}
	0 \\ 1
	\end{pmatrix}
	\notag \\	
	&=
	\frac{1}{\sqrt{N_\mathrm{cell}}}
	\sum_\mathbf{k} e^{i\mathbf{k}\cdot \mathbf{r}}
	\left(
	b_1 (\mathbf{k})|u_1(\mathbf{k})\rangle
	+
	b_2 (\mathbf{k})|u_2(\mathbf{k})\rangle
	\right).
\end{align}
We now prepare $|\mathrm{A}\rangle$ as an initial state, project the state onto the lower band, and consider time-evolution under shaking as discussed in the main text. Let us denote the projector to the lower band by $\mathcal{P}_1$. Then, the state after the projection is
\begin{align}
	\mathcal{P}_1|\mathrm{A}\rangle
	=
	\frac{1}{\sqrt{N_\mathrm{cell}}}
	\sum_\mathbf{k} e^{i\mathbf{k}\cdot \mathbf{r}}
	a_1 (\mathbf{k})|u_1(\mathbf{k})\rangle.
\end{align}
Consulting Eq.~(19) of the main text, we see that the integrated rate upon linear shaking along the $x$ direction, using $\mathcal{P}_1|\mathrm{A}\rangle$ as the initial state, is
\begin{align}
	\Gamma_{\mathrm{A}, \hat{x}}^{\mathrm{int}}
	=
	\frac{2\pi E^2}{\hbar}
	\frac{1}{N_\mathrm{cell}}\sum_\mathbf{k} |a_1 (\mathbf{k})|^2 g_{xx}(\mathbf{k}).
\end{align}	
Similarly, the integrated rate when we consider $\mathcal{P}_1|\mathrm{B}\rangle$ as the initial state is
\begin{align}
	\Gamma_{\mathrm{B}, \hat{x}}^{\mathrm{int}}
	=
	\frac{2\pi E^2}{\hbar}
	\frac{1}{N_\mathrm{cell}}\sum_\mathbf{k} |b_1 (\mathbf{k})|^2 g_{xx}(\mathbf{k}).
\end{align}
Since the matrix (\ref{unitarytransformation}) is a unitary matrix, the relation $|a_1(\mathbf{k})|^2 + |b_1(\mathbf{k})|^2 = 1$ holds. Then, the sum of the integrated excitation probability is
\begin{align}
	\Gamma_{\mathrm{A}, \hat{x}}^{\mathrm{int}}
	+
	\Gamma_{\mathrm{B}, \hat{x}}^{\mathrm{int}}
	=
	\frac{2\pi E^2}{\hbar}
	\frac{1}{N_\mathrm{cell}}\sum_\mathbf{k} g_{xx}(\mathbf{k}),
\end{align}
where $\sum_\mathbf{k} g_{xx}(\mathbf{k}) / N_\mathrm{cell}$ is nothing but the average of the $xx$-component of the quantum metric over the Brillouin zone.
Thus, the uniform filling of a band can be simulated by performing two set of measurements using states that are localized on single sites.


\begin{thebibliography}{99}

\bibitem{Mead_review} C. Alden Mead, {\it The geometric phase in molecular systems}, Rev. Mod. Phys. {\bf 64}, 51 (1992).

\bibitem{Xiao:2010} D. Xiao, M.-C. Chang, and Q. Niu, {\it Berry phase effects on electronic properties}, Rev. Mod. Phys. {\bf 82}, 1959 (2010).

\bibitem{Dalibard_gauge} J. Dalibard, F. Gerbier, G. Juzeliunas, and P. Ohberg, {\it Colloquium: Artificial gauge potentials for neutral atoms}, Rev. Mod. Phys. 83, 1523 (2011).

\bibitem{Nakahara}  M. Nakahara, {\it Geometry, Topology and Physics}, CRC Press; 2nd Edition (2003).  

\bibitem{Aharonov} Y. Aharonov and D. Bohm, {\it Significance of Electromagnetic Potentials in the Quantum Theory}, Phys. Rev. {\bf 115}, 485 (1959).

\bibitem{WuYang} T. T. Wu and C. N. Yang, {\it Concept of nonintegrable phase factors and global formulation of gauge fields}, Phys. Rev. D {\bf 12}, 3845 (1975).

\bibitem{Berry:1984} M. Berry, {\it Quantal phase factors accompanying adiabatic changes}, Proc. Roy. Soc. A {\bf 392}, 45 (1984).


\bibitem{Thouless:1982} D. Thouless, M. Kohmoto, M. Nightingale, and M. den Nijs, {\it Quantized Hall conductance in a two-dimensional periodic potential}, Phys. Rev. Lett. {\bf 49}, 405 (1982).

\bibitem{Hasan:2010} M. Z. Hasan and C. L. Kane, {\it Colloquium: Topological insulators}, Rev. Mod. Phys. {\bf 82}, 3045 (2010).

\bibitem{Qi:2011} X.-L. Qi and S.-C. Zhang, {\it Topological insulators and superconductors}, Rev. Mod. Phys. {\bf 83}, 1057 (2011).

\bibitem{Simon1983} B. Simon, {\it Holonomy, the Quantum Adiabatic Theorem, and Berry's Phase}, Phys. Rev. Lett. {\bf 51}, 2167 (1983).

\bibitem{Karplus} R. Karplus and J. M. Luttinger, {\it Hall Effect in Ferromagnetics}, Phys. Rev. {\bf 95}, 1154 (1954).

\bibitem{Aidelsburger_review} M. Aidelsburger, S. Nascimbene and N. Goldman, {\it Artificial gauge fields in materials and engineered systems}, arXiv:1710.00851 (2017).

\bibitem{Ozawa:2018} T. Ozawa, H. M. Price, A. Amo, N. Goldman, M. Hafezi, L. Lu, M. Rechtsman, D. Schuster, H. Simon, O. Zilberberg, and I.  Carusotto, {\it Topological Photonics}, arXiv:1802.04173.

\bibitem{Cooper:2018} N. R. Cooper, J. Dalibard, and I. B. Spielman, {\it Topological Bands for Ultracold Atoms}, arXiv:1803.00249.

\bibitem{Flaschner:2016} N. Fl\"aschner, B. S. Rem, M. Tarnowski, D. Vogel, D.-S. Lühmann, K. Sengstock, C. Weitenberg, {\it Experimental reconstruction of the Berry curvature in a topological Bloch band}, Science {\bf 352}, 1091 (2016).

\bibitem{Li:2016Science} T. Li, L. Duca, M. Reitter, F. Grusdt, E. Demler, M. Endres, M. Schleier-Smith, I. Bloch, and U. Schneider, {\it Bloch state tomography using Wilson lines}, Science {\bf 352}, 1094 (2016).

\bibitem{Bloch_AB} L. Duca, T. Li, M. Reitter, I. Bloch, M. Schleier-Smith, and U. Schneider, {\it An Aharonov-Bohm interferometer for determining Bloch band topology}, Science {\bf 347}, 288 (2015).

\bibitem{Wimmer:2017} M. Wimmer, H. M. Price, I. Carusotto, and U. Peschel, {\it Experimental measurement of the Berry curvature from anomalous transport}, Nat. Phys. {\bf 13}, 545 (2017).

\bibitem{Provost:1980} J. P. Provost and G. Vallee, {\it Riemannian structure on manifolds of quantum states}, Commun.Math. Phys. {\bf 76}, 289 (1980).

\bibitem{Anandan:1990} J. Anandan and Y. Aharonov, {\it Geometry of quantum evolution}, Phys. Rev. Lett. {\bf 65}, 1697 (1990).

\bibitem{Marzari:1997} N. Marzari and D. Vanderbilt, {\it Maximally localized generalized Wannier functions for composite energy bands}, Phys. Rev. B {\bf 56}, 12847 (1997).

\bibitem{Neupert:2013} T. Neupert, C. Chamon, and C. Mudry, {\it Measuring the quantum geometry of Bloch bands with current noise}, Phys. Rev. B {\bf 87}, 245103 (2013).

\bibitem{Kolodrubetz:2013} M. Kolodrubetz, V. Gritsev, and A. Polkovnikov, {\it Classifying and measuring geometry of a quantum ground state manifold}, Phys. Rev. B {\bf 88}, 064304 (2013).

\bibitem{Kolodrubetz:2017} M. Kolodrubetz, D. Sels, P. Mehta, and  A. Polkovnikov, {\it Geometry and non-adiabatic response in quantum and classical systems}, Phys. Rep. {\bf 697}, 1 (2017).

\bibitem{Albert:2016} V. V. Albert, B. Bradlyn, M. Fraas, and L. Jiang, {\it Geometry and response of Lindbladians}, Phys. Rev. X {\bf 6}, 041031 (2016).

\bibitem{Bleu:2016} O. Bleu, G. Malpuech, and D. D. Solnyshkov, {\it Effective theory of non-adiabatic quantum evolution based on the quantum geometric tensor}, arXiv:1612.02998.

\bibitem{Ozawa:2018PRB} T. Ozawa, {\it Steady-state Hall response and quantum geometry of driven-dissipative lattices}, Physical Review B {\bf 97}, 041108(R) (2018).

\bibitem{Raoux:2015} A. Raoux, F. Pi\'echon, J.-N. Fuchs, and G. Montambaux, {\it Orbital magnetism in coupled-bands models}, Phys. Rev. B {\bf 91}, 085120 (2015).

\bibitem{Gao:2015} Y. Gao, S. A. Yang, and Q. Niu, {\it Geometrical effects in orbital magnetic susceptibility}, Phys. Rev. B {\bf 91}, 214405 (2015).

\bibitem{Piechon:2016} F. Pi\'echon, A. Raoux, J.-N. Fuchs, and G. Montambaux, {\it Geometric orbital susceptibility: Quantum metric without Berry curvature}, Phys. Rev. B {\bf 94}, 134423 (2016).

\bibitem{Combes:2016} F. Combes, M. Trescher, F. Pi\'echon, and J.-N. Fuchs, {\it Statistical mechanics approach to the electric polarization and dielectric constant of band insulators}, Phys. Rev. B {\bf 94}, 155109 (2016). 

\bibitem{Freimuth:2017} F. Freimuth, S. Bl\"ugel, and Y. Mokrousov, {\it Geometrical contributions to the exchange constants: Free electrons with spin-orbit interaction}, Phys. Rev. B {\bf 95}, 184428 (2017).

\bibitem{Julku:2016} A. Julku, S. Peotta, T. I. Vanhala, D.-H. Kim, and P. T\"orm\"a, {\it Geometric origin of superfluidity in the Lieb-lattice flat band}, Phys. Rev. Lett. {\bf 117}, 045303 (2016).

\bibitem{Liang:2017} L. Liang, T. I. Vanhala, S. Peotta, T. Siro, A. Harju, and P. T\"orm\"a, {\it Band geometry, Berry curvature, and superfluid weight}, Phys. Rev. B {\bf 95}, 024515 (2017).

\bibitem{Iskin:2018} M. Iskin, {\it Quantum metric contribution to the pair mass in spin-orbit coupled Fermi superfluids}, arXiv:1801.09388.

\bibitem{CamposVenuti:2007} L. Campos Venuti and P. Zanardi, {\it Quantum critical scaling of the geometric tensors}, Phys. Rev. Lett. {\bf 99}, 095701 (2007).

\bibitem{Zanardi:2007} P. Zanardi, P. Giorda, and M. Cozzini, {\it Information-theoretic differential geometry of quantum phase transitions}, Phys. Rev. Lett. {\bf 99}, 100603 (2007).

\bibitem{You:2007} W.-L. You, Y.-W. Li, and S.-J. Gu, {\it Fidelity, dynamic structure factor, and susceptibility in critical phenomena}, Phys. Rev. E {\bf 76}, 022101 (2007).

\bibitem{Dey:2012} A. Dey, S. Mahapatra, P. Roy, and T. Sarkar, {\it Information geometry and quantum phase transitions in the Dicke model}, Phys. Rev. E {\bf 86}, 031137 (2012).

\bibitem{Ma:2010} Y.-Q. Ma, S. Chen, H. Fan, and W.-M. Liu, {\it Abelian and non-Abelian quantum geometric tensor}, Phys. Rev. B {\bf 81}, 245129 (2010).

\bibitem{Legner:2013} M. Legner and T. Neupert, {\it Relating the entanglement spectrum of noninteracting band insulators to their quantum geometry and topology}, Phys. Rev. B {\bf 88}, 115114 (2013).

\bibitem{Roy:2014} R. Roy, {\it Band geometry of fractional topological insulators}, Phys. Rev. B {\bf 90}, 165139 (2014).

\bibitem{Claassen:2015} M. Claassen, C. H. Lee, R. Thomale, X.-L. Qi, and T. P. Devereaux, {\it Position-momentum duality and fractional quantum Hall effect in Chern insulators}, Phys. Rev. Lett. {\bf 114}, 236802 (2015).

\bibitem{Bauer:2016} D. Bauer, T. S. Jackson, and R. Roy, {\it Quantum geometry and stability of the fractional quantum Hall effect in the Hofstadter model}, Phys. Rev. B {\bf 93}, 235133 (2016).

\bibitem{Lim:2015PRA} L.-K. Lim, J.-N. Fuchs, and G. Montambaux, {\it Geometry of Bloch states probed by St\"uckelberg interferometry}, Phys. Rev. A {\bf 92}, 063627 (2015). 

\bibitem{Srivastava:2015} A. Srivastava and A. Imamo\u{g}lu, {\it Signatures of Bloch-band geometry on excitons: Nonhydrogenic spectra in transition-metal dichalcogenides}, Phys. Rev. Lett. {\bf 115}, 166802 (2015).

\bibitem{Resta:2005} R. Resta, {\it Electron Localization in the Quantum Hall Regime}, Phys. Rev. Lett. {\bf 95}, 196805 (2005).

\bibitem{Thonhauser:2006PRB} T. Thonhauser and D. Vanderbilt, {\it Insulator/Chern-insulator transition in the Haldane model}, Phys. Rev. B {\bf 74}, R1651 (2006).

\bibitem{Marzari:2012} N. Marzari, A. A. Mostofi, J. R. Yates, I. Souza, and D. Vanderbilt, {\it Maximally localized Wannier functions: Theory and applications}, Rev. Mod. Phys. {\bf 84}, 1419 (2012).

\bibitem{Souza:2000} I. Souza, T. Wilkens, and R. M. Martin, {\it Polarization and localization in insulators: Generating function approach}, Phys. Rev. B {\bf 62}, 1666 (2000).

\bibitem{Souza_2008} I. Souza and D. Vanderbilt, {\it Dichroic 
f-sum rule and the orbital magnetization of crystals}, Phys. Rev. B {\bf 77}, 054438 (2008).

\bibitem{Tran:2017} D. T. Tran, A. Dauphin, A. G. Grushin, P. Zoller, and N. Goldman, {\it Probing topology by ``heating": Quantized circular dichroism in ultracold atoms}, Science Advances {\bf 3}, e1701207 (2017).

\bibitem{Juan:2017} F. de Juan, A. G. Grushin, T. Morimoto, J. E. Moore, {\it Quantized circular photogalvanic effect in Weyl semimetals}, Nature Comm. {\bf 8}, 15995 (2017).

\bibitem{Tran:2018} D. T. Tran, N. R. Cooper, and N. Goldman, {\it Quantized Rabi Oscillations and Circular Dichroism in Quantum Hall Systems}, arXiv:1803.01010.

\bibitem{Reitter2017} M. Reitter \emph{et al.}, {\it Interaction Dependent Heating and Atom Loss in a Periodically Driven Optical Lattice}, Phys. Rev. Lett. {\bf 119}, 200402 (2017).

\bibitem{Flaschner_2018} N. Fl\"aschner \emph{et al.}, {\it High precision spectroscopy of ultracold atoms in optical lattices}, arXiv:1801.05614.

\bibitem{Aidelsburger:2014} M. Aidelsburger, M. Lohse, C. Schweizer, M. Atala, J. T. Barreiro, S. Nascimb\`ene, N. R. Cooper, I. Bloch, and N. Goldman, {\it Measuring the Chern number of Hofstadter bands with ultracold bosonic atoms}, Nature Physics {\bf 11}, 162 (2015).

\bibitem{Harper:1955} P. G. Harper, {\it Single Band Motion of Conduction Electrons in a Uniform Magnetic Field}, Proc. Phys. Soc. A {\bf 68}, 874 (1955).

\bibitem{Hofstadter:1976} D. Hofstadter, {\it Energy levels and wave functions of Bloch electrons in rational and irrational magnetic fields}, Phys. Rev. B {\bf 14}, 2239 (1976).

\bibitem{Haldane:1998} F. D. M. Haldane, {\it Model for a Quantum Hall Effect without Landau Levels: Condensed-Matter Realization of the ``Parity Anomaly"}, Phys. Rev. Lett. {\bf 61}, 2015 (1988).

\bibitem{Roushan:2014Nature} P. Roushan, C. Neill, Y. Chen, M. Kolodrubetz, C. Quintana, N. Leung, M. Fang, R. Barends, B. Campbell, Z. Chen, B. Chiaro, A. Dunsworth, E. Jeffrey, J. Kelly, A. Megrant, J. Mutus, P. J. J. O'Malley, D. Sank, A. Vainsencher, J. Wenner, T. White, A. Polkovnikov, A. N. Cleland, and J. M. Martinis, {\it Observation of topological transitions in interacting quantum circuits}, Nature {\bf 515}, 241 (2014).

\bibitem{Lin:2009Nature} Y. J. Lin, R. L. Compton, K. Jim\'enez-Garc\'ia, J. V. Porto, and I. B. Spielman, Nature {\bf 462}, 628 (2009).

\bibitem{Jotzu:2015} G. Jotzu, M. Messer, R. Desbuquois, M. Lebrat, T. Uehlinger, D. Greif, and T. Esslinger, {\it Experimental realization of the topological Haldane model with ultracold fermions}, Nature {\bf 515}, 237 (2014).

\bibitem{Aidelsburger:2013} M. Aidelsburger, M. Atala, M. Lohse, J. T. Barreiro, B. Paredes, and I. Bloch, {\it Realization of the Hofstadter Hamiltonian with Ultracold Atoms in Optical Lattices}, Phys. Rev. Lett. {\bf 111}, 185301 (2013).

\bibitem{Miyake:2013} H. Miyake, G. A. Siviloglou, C. J. Kennedy, W. C. Burton, and W. Ketterle, {\it Realizing the Harper Hamiltonian with Laser-Assisted Tunneling in Optical Lattices}, Phys. Rev. Lett. {\bf 111}, 185302 (2013).

\bibitem{Kennedy:2015} C. J. Kennedy, W. C. Burton, W. C. Chung, and W. Ketterle, {\it Observation of Bose–Einstein condensation in a strong synthetic magnetic field}, Nature Physics {\bf 11}, 859 (2015).

\bibitem{Tarnowski:2017PRL} M. Tarnowski, M. Nuske, N. Fl\"aschner, B. Rem, D. Vogel, L. Freystatzky, K. Sengstock, L. Mathey, and C. Weitenberg, {\it Observation of topological Bloch-state defects and their merging transition}, Phys. Rev. Lett. {\bf 118}, 240403 (2017).

\bibitem{LandauBook} L. D. Landau and E. M. Lifshitz, {\it Quantum Mechanics: Non-Relativistic Theory, Volume 3} (Butterworth-Heinemann, 1981).

\bibitem{Sup} See Supplemental Material.

\end{thebibliography}
\end{document}